\documentclass[proceedings]{rmaa}

\title{TURBULENT DISSIPATION IN THE INTERSTELLAR MEDIUM: 
IMPLICATIONS FOR GALAXY FORMATION AND EVOLUTION}

\author{Vladimir Avila-Reese and Enrique V\'azquez-Semadeni
       \affil{Instituto de Astronom\'{\i}a, Universidad Nacional 
        Aut\'onoma de M\'exico}}

\fulladdresses{
\item Vladimir Avila-Reese and Enrique V\'azquez-Semadeni:
Instituto de Astronom\'{\i}a, UNAM, A.P. 70-268, 04510 M\'exico,
D.F., M\'exico (Emails: avila,enro@astroscu.unam.mx)}

\shortauthor{Avila-Reese \& V\'azquez-Semadeni}
\shorttitle{Turbulent dissipation in the ISM}


\resumen{Estudiamos la disipaci\'on turbulenta en el MIE y 
algunas implicaciones para la formaci\'on y evoluci\'on de 
galaxias usando simulaciones MHD num\'ericas de fluidos 
compresibles en 2D. La energ\'{\i}a cin\'etica $E_k$ se inyecta 
por fuentes estelares formadas autoconsistentemente. En el fluido
coexisten el r\'egimen de turbulencia forzada y el de decaimiento.
En la regiones turbulentas activas, $E_k$ es 
disipada local y eficientemente. En el 
r\'egimen de decaimiento $E_k(t)$ decae 
$\sim (1+t)^{-0.8 }$. Los movimientos turbulentos residuales pueden
propagarse distancias del orden del alto del disco gaseoso
lo cual sugiere que la turbulencia puede propiciar el 
soporte vertical con una formaci\'on estelar autoregulada al nivel
del disco, mas no al nivel de todo el halo cosmol\'ogico como se
requiere en ciertos modelos de formaci\'on de galaxias.}

\abstract{We study turbulent dissipation in the ISM and explore
some implications for galaxy formation and evolution using 2D 
MHD numerical simulations of compressible fluids. The
turbulent kinetic energy $E_k$ is injected by stellar sources 
formed self-consistently in the simulation. In the ISM-like fluid,
regimes of both forced and decaying turbulence coexist. In the active
turbulent regions (forced regime), $E_k$ is dissipated locally and
efficiently.
In the decaying regime (far from input sources), $E_k(t)$ 
decays $\sim (1+t)^{-0.8 }$. The residual turbulent motions may 
propagate distances of the order of the observed disk height, 
suggesting that turbulence may be the
responsible of vertical support and star formation self-regulation 
at the disk level, but not at the level of the whole cosmological
halo, as would be required in some models of galaxy
formation.}

\nonstopmode

\begin{document}
\maketitle

\section{Introduction}

Modeling galaxy formation and evolution requires 
of a solid cosmological theoretical framework as much as an adequate
model to describe the large-scale star formation (SF) cycle and its
interplay with the interstellar medium (ISM). The dissipative
properties of the ISM play a crucial role in this latter process.
Globally, stellar radiation is mainly responsible for maintaining
the temperature of the ISM in its various phases.
However, thermal pressure is thought to be 
negligible for the global disk gas dynamics (e.g., V\'azquez-
Semadeni et al. 1999; Cox, these proceedings).  
Nevertheless, stars are also sources of kinetic energy ($E_k$) 
deposition into the ISM, and as hydrodynamical 
simulations have shown, the dynamics of the gas in this case is 
deeply affected (e.g., Navarro \& White 1993). Due to the large $E_k$ 
input and the high Reynolds number of the ISM plasma, turbulence is 
expected to develop, its pressure and dissipation 
being key ingredients in the ``metabolism'' of the 
disk stellar-gas system.

Several disk galaxy evolution models (e.g., Firmani, Hern\'andez,
\& Gallagher 1996) are based on the idea that
the intrinsic SF rate (SFR) is controlled by a balance 
within the vertical {\it disk} gas between the turbulent energy input 
rate due to SF and the dissipation rate. The crucial parameter 
for the SFR and disk height is the turbulent dissipation timescale. 

The self-regulating SF mechanism has been also used in  
models of galaxy formation within the context of the 
hierarchical CDM-based scenario, but in this case it was
applied to the large cosmological halo (White 
\& Frenk 1991; Kauffmann, White, \& Guiderdoni 1993; Cole et 
al. 1994; Somerville \& Primack 1999; van den Bosch 1999). 
In these models the feedback of the stars is assumed to 
efficiently reheat and drive back the disk gas into the 
dark matter halo, in such a
way that the SFR efficiency is a strong function of the
halo mass. Thus, a crucial question is whether the energy 
released by SNe and stars is able to not only maintain
the warm and hot phases and the stirring of the ISM, but 
also to sustain a huge hot corona in quasi-hydrostatic
equilibrium with the cosmological halo. 

It should be emphasized that the observed medium around
the disks ---diffuse ionized and 
high-velocity-dispersion HI gas, usually called the 
halo--- is much more local than the hypothetical
gas in virial equilibrium with the huge dark halo. We shall
refer to the former as the the {\bf extraplanar medium}, and 
to the latter as the {\bf intrahalo medium}.
It is still not at all clear how to explain the observed 
extraplanar medium, particularly the ionized gas (see a 
recent review by Mac Low 1999). This calls into 
question the possibility that ionizing sources from the 
disk (mainly massive OB stars) are able to sustain the 
extended intrahalo medium. A possibility is that this gas 
is heated by turbulence from the disk. However, this 
question again depends on the ability of the turbulent 
ISM to dissipate its $E_k$. Avila-Reese \& V\'azquez-Semadeni 
(in preparation; hereafter AV) have studied the 
dissipative properties of compressible MHD fluids that 
resemble the ISM. Here we briefly report their main results 
and remark the implications on the aforementioned questions.
  
\section{The method}

AV have used numerical 2D MHD simulations of self-gravitating
turbulent compressible fluids that include terms for radiative
cooling, heating, rotation and stellar energy injection 
(V\'azquez-Semadeni, Passot, \& Pouquet 1995,1996; Passot, 
V\'azquez-Semadeni, \& Pouquet 1996). The parameters were chosen 
in such a way the simulations resemble the ISM in the plane of 
the Galaxy at the {\bf 1 kpc scale}. 

Previous simulations on dissipation in compressible MHD fluids 
were focused to study molecular clouds (Mac Low et al. 
1998; Stone, Ostriker, \& Gammie 1998; Padoan \& Nordlund 1999; 
Mac Low 1999). In those works where the forced case was
studied, the turbulence was driven in Fourier space by large-scale 
random velocity perturbations whose 
amplitudes are selected as to maintain $E_k$ constant in time. 
As a result, $E_k$ is injected everywhere in space 
(a ``ubiquitous'' injection). Instead, in the ISM the 
stellar input sources are pointlike and their spheres of 
direct influence are comparatively small w.r.t. to typical 
scales of the global ISM. In the simulations of AV, an 
``energy input source'' is turned on at grid point $\vec x$ 
whenever $\rho (\vec x)>\rho _c$, and $\vec \nabla \cdot 
\vec u(\vec x)<0$. Once SF has turned on at a given grid point, 
it stays on for a time interval $\Delta t_s$, during
which the gas receives an 
acceleration $\vec a$ directed radially away from this point.
The input sources are spatially extended by convolving their spatial 
distribution with a Gaussian of width $\lambda _f$. For
the turbulent fluid, $\lambda _f$ is the forcing scale. At 
this scale the acceleration $\vec a$ produces a velocity difference
around the ``star'' $\upsilon _f\approx 2\vec {a}\Delta t_s$ (the 
velocity at which turbulence is forced at the $\lambda _f$ scale).
Both $\lambda _f$ and $\upsilon _f$ are free parameters.

\section{Dissipation in driven and decaying regimes}

For ISM simulations ($128^2$) with driven 
turbulence, AV find that the behaviour with time of the $E_k$ 
dissipation rate, $E_k^d$, is similar to that of the $E_k$ injection 
rate, $\dot{E}_k$. This means that $E_k$ is dissipated locally, near
the input sources. For various simulations varying $\lambda _f$ and 
$\upsilon _f$, it was found that the the dissipation timescale is given by
$t_d\approx 1.5-3.0\ 10^7(\frac{\lambda _f/30 pc}{\upsilon _f/30kms^{-1}})$,
i.e., $t_d$ is proportional to $\lambda _f/\upsilon _f$. 

Due to the locality and discretness of the energy input 
sources, most of the volume actually is occupied by a turbulent 
flow in a decaying regime. Thus, one may say that in the same fluid
``active'' turbulent regions, where the turbulence
driven by small non-ubiquitous input sources 
is {\it locally} dissipated, coexist with extended regions with a 
``residual'' turbulence in decaying regime and
characterized by $\upsilon _{\rm rms}$, where 
$\upsilon _{\rm rms}\ll \upsilon _f$.
In order to study the decaying regime, SF was turned off
in the simulations after some time $t\gg t_{d}$. It was found
that $E_k$ decays as $(1+t)^{-n}$ with $n\sim 0.8$, in good 
agreement with previous studies for isothermal fluids 
(Mac Low et al. 1998, Stone et al. 1998). A typical 
decaying timescale, $t_{\rm dec}$, may be defined as the time at which 
the initial $E_k$ has decreased by a factor 2.
From our simulations, $t_{\rm dec}\approx 1.7\times 10^7$ years,
which is in agreement with $t_d$ in the driven turbulence. With these
timescales, ``residual'' turbulent motions  propagating at roughly 10 km/s
would attain typical distances  of approximately 200 pc. It was also
suggested, on dimmensional arguments, that $E_k$
and $\upsilon _{\rm rms}$ will decay with distance $\ell$
as $\ell^{-2m}$ and $\ell^{-m }$, respectively, with $m=n/(2-n)$.

\section{Conclusions and implications}

$\bullet$ Localized, discrete forcing at small
scales gives rise to the coexistence of both forced and 
decaying turbulence regimes in the same flow (ISM).

$\bullet$ The turbulent $E_k$ near the ``active'' turbulent 
regions is dissipated locally and efficiently. The global dissipation 
timescale $t_d$ is proportional to $\lambda _f/\upsilon _f$. For reasonable
values of $\lambda _f$ and $\upsilon _f$ (which produce $\upsilon _{\rm
rms}\sim 10$ km/s), $t_d$ is of the order of a few $10^7$ years. Far from
the sources, the ``residual'' ISM turbulence  decays as $E_k(t)\propto
(1+t)^{-0.8}$. The characteristic decay time is again a few $10^7$
years, and for $\upsilon _{\rm rms}\sim 10$ km/s, the
turbulent motions reach distances of $\sim 200$ pc.

$\bullet$ Turbulent motions produced in
the disk plane will propagate up to distances of the order of
the gaseous disk height. Therefore, models of
galaxy evolution where this height is 
determined by an energy balance that self-regulates 
SF in the ISM appear viable. However, our results pose a serious
difficulty for models of galaxy
formation where the turbulent $E_k$ injected by SNe is
thought to be able to reheat and drive back the gas from the disk
into the intrahalo medium in such a way that the
SF is self-regulated at the level of the cosmological halo.
Nevertheless, for non-stationary runaway SF 
(starbursts), most of the superbubbles might be able to 
blowout of the disk, as required for expelling large amounts 
of gas and energy into the dark matter halo.

\end{document}